# Prediction of Individual Propofol Requirements based on Preoperative EEG Signals

Young-Seok Kweon[1], Minji Lee[1], Dong-Ok Won[1], Kwang-Suk Seo[2]
[1]Department of Brain and Cognitive Engineering, Korea University, Seoul, Republic of Korea
[2]Department of Dental Anesthesiology, Seoul National University, Seoul, Republic of Korea
youngseokkweon@korea.ac.kr, minjilee@korea.ac.kr, wondongok@korea.ac.kr, stone90@snu.ac.kr

*Abstract*—The patient must be given an adequate amount of propofol for safe surgery since overcapacity and low capacity cause accidents. However, the sensitivity of propofol varies from patient to patient, making it very difficult to determine the propofol requirements for anesthesia. This paper aims to propose a neurophysiological predictor of propofol requirements based on the preoperative electroencephalogram (EEG). We exploited the canonical correlation analysis that infers the amount of information on the propofol requirements. The results showed that the preoperative EEG included the factor that could explain the propofol requirements. Specifically, the frontal and posterior regions had crucial information on the propofol requirements. Moreover, there was a significantly different power in the frontal and posterior regions between baseline and unconsciousness periods, unlike the alpha power in the central region. These findings showed the potential that preoperative EEG could predict the propofol requirements.

## I. Introduction

Propofol is one of the most commonly used anesthetics in surgical environments in that it can cause unresponsive such as non-rapid eye movement sleep [1-3]. Rapid recovery from anesthesia and ease to reduce pain are the reason why propofol is widely used [4]. However, since the sensitivity to propofol varies from person to person [5] and from mental state to mental state [6], it is difficult to estimate the dosage of propofol that switches from consciousness to unconsciousness and over-capacity or low capacity of propofol happen. The overcapacity of propofol can cause secondary accidents such as airway closure and respiratory failure [7], and the low capacity of propofol causes another accident, such as states of awareness during anesthesia [8]. So, determining the propofol requirements that switches from consciousness to unconsciousness is a crucial issue.

Several types of research have proposed predictors to determine the propofol requirements [6, 9-10]. Kil et al. [6] proposed that preoperative anxiety from Spielberger's state-trait anxiety inventory would be a predictor of propofol requirements. They showed the coefficient of determination, $R^2 = 0.127$ between preoperative anxiety and propofol requirements. Masuda et al. [9] also operated questionnaires but focused on patient satisfaction to determine optimal propofol requirements. Since they were based on questionnaires, it is difficult to understand what kind of brain activity associated with propofol requirements. Moreover, Araújo et al. [10] exploited body mass index (BMI) to predict the propofol requirements, which was the blood concentration of propofol (CE) at loss of consciousness. They only suggest a predictor for patients without obesity using body mass index. Previous research did not consider brain activity. Ignoring the brain activity seemed to be unreasonable when predicting propofol requirements because propofol affects brain activity that could be measured by EEG or other equipment.

EEG is one of the methods to measure brain activity [11-15]. Unlike other methods such as functional magnetic resonance imaging (fMRI) and functional near-infrared spectroscopy (fNIRS), which obtain the hemodynamic signal, EEG measures differences of electrical potential on the scalp [16]. So, fMRI and fNIRS have a poor temporal resolution because the hemodynamic response has a delay from initial brain activity [17]. The high temporal resolution, low cost, and spending small space are the advantages of EEG [18]. Since surgical environments are complicated, portable and low-costed equipment would be beneficial for surgery. Moreover, high temporal resolution is a suitable characteristic because we must consider instant changes in brain activity, such as motor imagery [19-20]. The changes in EEG characteristics during the unconsciousness period are increasing low-frequency power and decreasing alpha coherence in the occipital region [21]. Distinguishing individual EEG differences from common EEG differences like Kim et al. [22] during anesthesia help to make the neurophysiological predictor for the propofol requirements.

Patient-controlled sedation (PCS) is the adaptation of patient-controlled drug administration techniques to the supply of intra-operative sedation [23]. Since administration is not constant and is operated by patients' own will, the EEG changes during the CE dynamics can be investigated. In PCS, there can be multiple loss and recovery of consciousness. Among them, the first loss of consciousness was the point from wakefulness to unconsciousness. The CE at the loss of consciousness could represent the propofol requirements for loss of consciousness [24]. So, we set that the propofol requirements are the CE at the first loss of consciousness.

In this study, we tried to validate that the preoperative EEG has the information to predict propofol requirements. Moreover,

This work was supported by Institute for Information & Communications Technology Promotion (IITP) grant funded by the Korea government (No. 2017-0-00451, Development of BCI based Brain and Cognitive Computing Technology for Recognizing User's Intentions using Deep Learning).

we divided brain regions into three regions to verify that what regions contributed to the predictor and changed between baseline and unconsciousness period. Our findings would help to predict the appropriate dosage of propofol required for safe surgery.

## II. METHODS

### A. Experimental Paradigm

The dataset has been published in Yeom et al. [25]. The thirty subjects (27.63 ± 7.06 years; F = 6) participated in this study. They had no history of neurological disorders. This study also was approved by the Institutional Review Board (IRB) at Seoul National University Dental Hospital (KCT0001618).

The subjects closed their eyes for 5 min before the propofol administration and measured EEG. They were instructed to press the button when they listened to the auditory stimuli ('Press the button'). Stimulation was presented with 9-11 sec intervals. Based on the dosage of propofol, subjects divided into three groups: high, medium, and low propofol groups. Each group was administered with 0.5, 0.3, 0.1 mg/kg, respectively. At the baseline session, propofol was not administered, although pressing the button. Otherwise, when they pressed a button at the resting session, propofol was injected with the lock-out time. The lock-out time of injection was set to 3 min for high propofol group and 1 min for medium and low propofol groups to prevent over-administration of propofol. As the concentration of propofol increased, the subjects switched from consciousness to unconsciousness. They did not press the button during propofol-induced unconsciousness even if auditory stimulation was presented. Through this PCS method with behavior tasks, we were able to explore the specific transition points from consciousness to unconsciousness.

Under the assumption that the subject followed the instruction thoroughly, we defined the loss of consciousness and the recovery of consciousness. The situation that subjects did not press the button with consecutive five times is the loss of consciousness [11]. Oppositely, the first point that the subject pressed the button after being unconscious was defined as the recovery of consciousness.

### B. Data Acquisition and Preprocessing

EEG data were measured from 62 Ag/AgCl electrodes based on the international 10−20 system using the BrainAmp EEG amplifier (Brain Products GmbH, Munich, Germany). The sampling rate was 1kHz, and we resampled the signal to 250 Hz. We performed a bandpass filter from 0 to 45 Hz. The data processing was performed with the BBCI toolbox based on MATLAB (The MathWorks, USA).

The signals from the baseline session were segmented for non-overlapping 8 sec after auditory stimulation was presented. These signals divided into four epochs for non-overlapping 2 sec. There was a total of 30 auditory stimulation for 5 min. Therefore, we obtained the 120 trials in the baseline session. The resting session consisted of the signals at consciousness and unconsciousness periods. Based on behavior tasks, we set the loss and recovery of consciousness. At the first loss of consciousness point in the resting session, segmentation that was the same process in the baseline session was performed to obtain the value for the unconsciousness period, which represents epochs for non-overlapping 2 sec. From this data, a periodogram that is an estimate of the spectral density of a signal was obtained. We calculated the power spectral density (PSD) from the periodogram at each channel and trials. The trial-average was performed to obtain subjects' PSD of each channel by baseline and unconsciousness. Moreover, we exploited 4 frequency bands: delta (0.5-4 Hz), theta (4-8 Hz), alpha (8-12 Hz), and spindle (12-16 Hz) bands. The PSD was integrated between each frequency band. We compared the delta, theta, alpha, and spindle PSD in frontal, central, posterior, and all regions between the baseline and unconsciousness with a paired t-test. It was performed to verify that the features were related to consciousness. Significance was assessed at the statistical level, $\alpha = 0.050$.

### C. Canonical Correlation Analysis

Since the EEG channel has the volume conduction property [26], it is crucial to fuse information from the different channels for a more accurate prediction of the propofol requirements. The canonical correlation analysis (CCA) seeks a set of linear transforms that project the features of different channels to a common space so that they can be comparable [27]. When there are two column vectors $X = (x_1, ..., x_n)^T$ and $Y = (y_1, ..., y_n)^T$ where $x_i \in \mathbb{R}^d$ and $y_i \in \mathbb{R}$, CCA search vectors $a \in \mathbb{R}^d$ and $b \in \mathbb{R}$ such that the $a^T X$ and $b^T Y$ maximize the correlation between the $a^T X$ and $b^T Y$.

$$(a', b') = argmax_{(a,b)} tr(a^T \Sigma b), \ \Sigma = XY^T \quad (1)$$

where $\Sigma$ denotes a multi-channel feature matrix's covariance matrix, and $tr(\cdot)$ represents trace function, which is the sum of diagonal elements.

In this study, we performed a CCA with propofol requirement from the PSD of 62 channels. We set that $X$ is the PSD of 62 channels from 30 subjects, which means $x_i \in \mathbb{R}^{62}$. Also, $Y$ was established as the propofol requirements. The projections of the original features onto their respective canonical bases can be considered as canonical presentation. The following equation represented the canonical presentation.

$$C = (a')^T X \quad (2)$$

The canonical presentation, $C$, is an unknown factor, which can explain the propofol requirements, but we could not directly measure using EEG. Since the canonical presentation do not represent the propofol requirements, it converted to predicted CE value by using linear regression. For testing the prediction performance, we obtained root mean squared error between predicted CE value and the propofol requirements.

We divided the 62 channels into three brain regions that are the frontal, central, and posterior regions to investigate which regions are essential for propofol requirements. The frontal region contains the 13 channels that are FP1, FP2, AF3, AF1, AF2, AF4, F7, F3, F1, Fz, F2, F4, and F8. The 30 channels that are FT9, FT7, FC5, FC3, FC1, FC2, FC4, FC6, FT8, FT10, T7,

C5, C3, C1, Cz, C2, C4, C6, T8, TP9, TP7, CP5, CP3, CP1, CPz, CP2, CP4, CP6, TP8, and TP10 were defined as the central region. We defined that the posterior region was the set of 19 channels, which are P7, P5, P3, P1, Pz, P2, P4, P6, P8, PO7, PO3, POz, PO4, PO8, PO9, O1, Oz, O2, and PO10. Then, we performed CCA using each region and analyzed the change of PSD between baseline and unconscious.

*D. Evaluation of Predicted CE*

We exploited the leave-one-out method to test the CCA results. Using 29 subjects from 30 subjects, CCA was performed. One subject excluded to test that results of CCA can predict the propofol requirements. The root means squared error (RMSE) was obtained to evaluate how similar propofol requirements were predicted by the CCA. We calculated the RMSE between the prediction of CCA and the propofol requirements.

## III. RESULTS

*A. Canonical Representation for Propofol Requirements*

We investigated the root mean squared error from each brain region. When using all channels, we obtained RMSEs that were 5.486, 4.500, 3.211, and 3.684 in the delta, theta, alpha, and spindle bands, respectively. Also, RMSEs were 4.317, 8.718, 7.668, and 13.895 when using the central region in the delta, theta, alpha, and spindle bands, respectively. However, we showed a lower RMSEs that were 2.521, 3.242, 2.283, and 2.459 when using the posterior regions in the delta, theta, alpha, and spindle bands, respectively. The frontal region's PSD showed the low RMSEs that were 1.367, 1.830, 1.921, and 1.591 in the delta, theta, alpha, and spindle bands, respectively (Table I). The RMSE when using PSD of frontal regions was lower than when using PSD of central and posterior PSDs. The lowest RMSE was shown in the frontal region when using the delta frequency band.

*B. Differences of PSD during State Changes*

Figure 1 showed that there were significant increments in delta PSDs from the frontal, central, posterior, all regions between baseline and unconsciousness (frontal: $p = 0.039$, central: $p < 0.001$, posterior: $p < 0.001$, all: $p < 0.001$). The theta PSDs in the frontal, central, posterior, all regions significantly increased (frontal: $p < 0.001$, central: $p = 0.001$, posterior: $p < 0.024$, all: $p = 0.003$). The alpha PSD also increased from the frontal region ($p = 0.013$). On the other hand, the alpha PSD in the posterior significantly decreased ($p < 0.001$). We showed no significantly different alpha PSDs in the central and all regions between baseline and unconsciousness (central: $p = 0.715$, all: $p = 0.286$). There were the spindle PSDs in the frontal, central, posterior, and all regions between baseline and unconsciousness (frontal: $p < 0.001$, central: $p < 0.001$, posterior: $p < 0.001$, all: $p < 0.001$).

TABLE I. ROOT MEAN SQUARE ERROR FROM EACH REGION

|         | *Frontal* | *Central* | *Posterior* | *All* |
|---------|-----------|-----------|-------------|-------|
| Delta   | 1.367     | 4.317     | 2.521       | 5.486 |
| Theta   | 1.830     | 8.718     | 3.242       | 4.500 |
| Alpha   | 1.921     | 7.668     | 2.283       | 3.211 |
| Spindle | 1.591     | 13.895    | 2.459       | 3.684 |

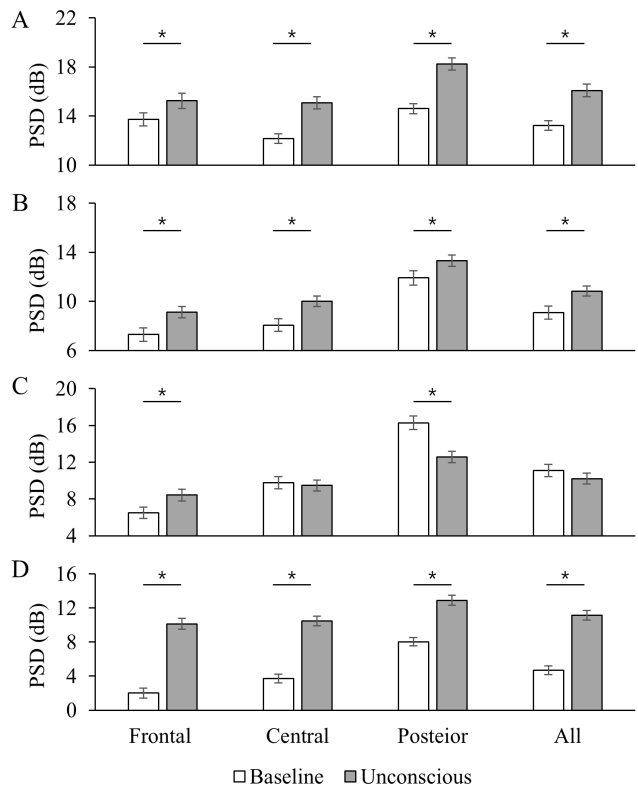

Figure 1. Comparison of PSD changes according to brain regions between baseline and unconscious states in (A) delta, (B) theta, (C) alpha, (D) spindle frequency bands. * indicates significant differences of PSD based on paired t-test. The error bar represents standard error.

## IV. DISCUSSION

In this study, we investigated the information about propofol requirements on preoperative EEG signals. The linear combination of canonical correlation using all channels' PSD showed that the canonical representation, which was obtained by CCA, could predict the propofol requirements. When we only exploited the fontal PSD, the prediction performance of propofol requirements was better than when utilizing the central and posterior regions. Moreover, we explored the changes in each region's PSD from baseline to unconsciousness period to verify that they related to consciousness. We found that the alpha PSD when using central and all regions was not changed between the baseline and unconsciousness, unlike the other PSD. Since alpha PSDs in central and all regions were not related to consciousness, it was not surprising that prediction performance using alpha PSD in the central and all regions was low. Although we showed that spindle and theta PSD in the central region significantly increased, the prediction performance using spindle and theta PSD in the central region was low. It might be because increments of spindle and theta PSD in the central region have small individual variance, which meant that all subjects show similar changes and original values. The delta PSD in the frontal region at the baseline session showed the highest performance and significantly changed between baseline and unconscious. Therefore, we concluded that delta PSD in the frontal region was a major component of the prediction of propofol requirements.

Our comparison of each regions' PSD coincided with previous research. We showed that there were significant increments of the delta, theta, alpha, and spindle PSD in the frontal region. Purdon et al. [21] showed that the power in low frequency and alpha band at the frontal region increased. Moreover, Xi et al. [28] showed that the power in the delta, theta, alpha, and spindle bands increased in the frontal region. In this study, we verified that the significant decrement of alpha PSD in the posterior region and constancy of alpha PSD in the central region. Gugino et al. [29] reported that there was no significant change of alpha PSD in the central region and the decrement of alpha PSD in the posterior region when becoming unconscious. The results that spindle PSD increased in all regions coincided with Kishimoto et al. [30]. Therefore, our results of the PSD comparison validated previous works' results.

The findings of our study showed that the PSD in the frontal and posterior regions had the information to predict the propofol requirements. The frontal region is a well-known region to operate high-level cognitive functions such as emotional expression, problem-solving, and memorizing [31]. When people lost consciousness, they also lost high-level cognitive function [21]. They could not remember during unconsciousness. Therefore, a significant change of PSD in the frontal region seemed reasonable. The posterior region, which was called as posterior hot zone was recently focused by neuroscientist [32]. They claimed that the posterior region contributed to the conscious experience than the frontal region. Patients in unconsciousness could not experience the outside world. So, changing the brain activity in the posterior region might have occurred because the posterior region's activity means each experience from the outside world.

Although our results suggested the implication of consciousness and strength compared to previous research, there were a few limitations. First, our experimental setting included behavior tasks and sensory processing in the baseline. We will need to verify that there were the same results when patients did not do and listen to anything. Second, we needed more analysis, such as spectral analysis and brain connectivity. Previous research showed that specific frequency bands changed differently and have focused on brain connectivity like the weighted phase lag index [2], Granger causality [33], and dynamic functional connectivity [34]. These can be the candidates of neurophysiological predictors for propofol requirements. Third, we exploited wet electrodes, but it is too inconvenient to utilize for the real-life surgical environment. So, we should verify that it is possible using dry electrodes [35]. Finally, we could not guarantee it could explain the new subject. So, we need to test our method for more subjects and other surgical environments, such as target-concentration infusion.

In summary, we investigated the information about propofol requirements on preoperative EEG signals. We verified that there was information about propofol requirements on preoperative EEG signals using CCA. Especially, delta PSD in the frontal region showed the highest performance of prediction.

## V. CONCLUSION AND FUTURE WORKS

We reported that the delta PSD in the frontal region has information for propofol requirements using the CCA. Also, we showed that the delta PSD in the frontal region was directly related to consciousness. Our results could contribute to help in to predict the propofol requirements. In the future, we will investigate other features, such as functional connectivity and entropy, to predict the propofol requirements. The various regression methods will be performed to improve prediction performance. Moreover, we will propose the predictor that can predict the various anesthetic requirements. Our future works will contribute to predict the anesthetic requirements.